\documentclass[aps,pra,superscriptaddress,floatfix,showpacs,10pt, twocolumn]{revtex4}
\usepackage[dvips]{graphicx}
\usepackage{amsmath}

\begin{document}

\title{Scheme for generating the cluster states via atomic ensembles}
\author{Ping Dong}
\email{pingdong@ahu.edu.cn}
\author{Zheng-Yuan Xue}
\email{zyxue@ahu.edu.cn}
\author{Ming Yang}
\email{mingyang@ahu.edu.cn}
\author{Zhuo-Liang Cao}
\email{zlcao@ahu.edu.cn}

\affiliation{School of Physics {\&} Material Science, Anhui
University, Hefei, 230039, P R China}

\begin{abstract}
A scheme for generating the cluster states via atomic ensembles is
proposed. The scheme has inherent fault tolerance function and is
robust to realistic noise and imperfections. All the facilities
used in our scheme are well within the current technology.
\end{abstract}

\maketitle

Entanglement is one of the most counterintuitive features in
quantum mechanics. One can realize many impossible tasks within
the classical world with the help of entangled states. Some
striking applications of entanglement including quantum dense
coding \cite{dense}, quantum teleportation \cite{teleport} and
quantum cryptography \cite{ekert} have been proposed. While
bipartite entangled state is well understood, multipartite
entanglement is still under extensive exploration. People soon
realized that it isn't just an extension of bipartite
entanglement. It not only inviolates local realism in a much
stronger way, but also is a useful resource in quantum information
processing. For tripartite entangled quantum system, it falls into
two classes of irreducible entanglement
\cite{trientangle,Vidal,Bru}. Now great efforts are engaged in the
investigation of multipartite entanglement with its promising
features, such as, decoherence-free quantum information
processing, multiparty quantum communications and so on. Recently,
Briegel \emph{et al}. \cite{briegel} introduced a class of
\emph{N}-qubit entangled states, \emph{i.e}., the cluster states,
which share the entanglement properties both in the GHZ- and
W-class entangled states. But they still have some different
properties, \emph{e.g}., they (in the case of N>4) are harder to
be destroyed by local operations than GHZ-class states. In
addition, they are also an universal resource for quantum
computation \cite{compute}.

Due to its promising applications in quantum information
processing and quantum computation, the cluster states have
attracted much attention. Recently Zou \emph{et al}. proposed
probabilistic schemes for generating the entangled cluster states
of four distant trapped atoms in leaky cavities \cite{zou1}, of
atoms in resonant microwave cavities \cite{zou2} and of
four-photon polarization via linear optics \cite{zou3}. Meanwhile,
many researchers pay their attentions to atomic ensembles in
realizing the scalable long-distance quantum communication
\cite{duan1}. The schemes based on atomic ensembles have some
peculiar advantages compared with the schemes of quantum
information processing by the control of single particles.
Firstly, the schemes have inherent fault tolerance function and
are robust to realistic noise and imperfections. Laser
manipulation of atomic ensembles without separately addressing the
individual atoms is dominantly easier than the coherent control of
single particles. In addition, atomic ensembles with suitable
level structure could have some kinds of collectively enhanced
coupling to certain optical mode due to the multi-atom
interference effects. Due to the above distinct advantages, a lot
of novel schemes for the generation of quantum entangled states
and quantum information processing have been proposed by using
atomic ensembles \cite{duan2,duan3,xue,lukin,liu}.

\begin{figure}[tbp]
\includegraphics[scale=0.35,angle=0]{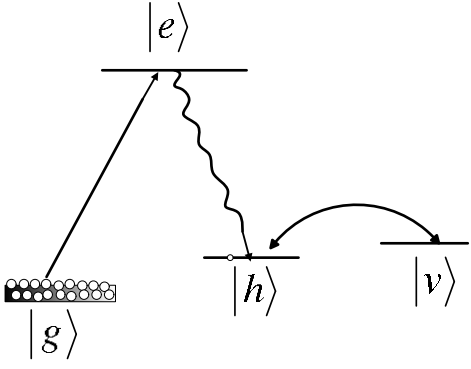}
\caption{The relevant atomic level structure of alkali metal atom.
The transition of $|e\rangle\rightarrow|h\rangle$ can emit a
forward-scattered Stokes photon that is co-propagating with the
laser pulse. The excitation in the mode $h$ can be transferred to
optical excitation by applying an anti-pump pulse.} \label{fig1}
\end{figure}

Here, we suggest a scheme to generate the cluster states via
atomic ensembles with a large number of identical alkali metal
atoms as basic system. The relevant level structure of the alkali
metal atoms is shown in Fig. \ref{fig1}. $|g\rangle$ is the ground
state, $|e\rangle$ is the excited state and $|h\rangle$,
$|v\rangle$ are two metastable states for storing a qubit of
information, \emph{e.g}., Zeeman or hyperfine sublevels. For the
three levels $|g\rangle$, $|h\rangle $ and $|v\rangle $, which can
be coupled via a Raman process, two collective atomic operators
can be defined as
$$s=(1/\sqrt{N})\Sigma_{i=1}^{N_{a}}|g\rangle_{i}\langle s|,$$
where $s = h,v$, and $N_{a}\gg1$ is the total number of atoms. $s$
is similar to independent bosonic mode operators providing all the
atoms remain in ground state $|g\rangle$. The states of the atomic
ensemble can be express as $|s\rangle = s^{+}|vac\rangle$ ($s =
h,v$) after the emission of the single Stokes photon in a forward
direction, where $|vac \rangle \equiv \otimes
_{i=1}^{N_{a}}|g\rangle _{i}$ denotes the ground state of the
atomic ensemble.

We firstly briefly discuss the generation of bipartite cluster
states via atomic ensembles. The atomic ensembles 1 and 2 are both
initially prepared in the state
\begin{equation}
\label{1} |\phi\rangle_{12} =v^{+}_{1}v^{+}_{2}|vac\rangle_{12}
\end{equation}
using Raman pulses. All the single-qubit transformation can be
achieved by laser pulses in atomic ensembles. Secondly, we perform
a single-qubit operation on atomic ensemble 1
\begin{equation}
\label{2} v^{+}_{1}|vac\rangle_{1} \rightarrow (h^{+}_{1}+
v^{+}_{1})|vac\rangle_{1}/\sqrt{2}.
\end{equation}
Then, we perform a controlled-not transformation on the two atomic
ensembles, where atomic ensemble 1 serving as controlled qubit and
atomic ensemble  2 as target qubit. The scheme for implementing
controlled-not gate via atomic ensembles with the help of Raman
laser manipulations, beam splitters, and single-photon detections
have been proposed in Ref. \cite{ping}. Now, the above procedures
lead Eq. (\ref{1}) to
\begin{equation}
\label{3} |\phi\rangle_{12} = (h^{+}_{1}v^{+}_{2}+
v^{+}_{1}h^{+}_{2})|vac\rangle_{12}/\sqrt{2}.
\end{equation}
Finally, we perform a single-qubit operation on atomic ensemble 1
\begin{eqnarray}
\label{4} h^{+}_{1}|vac\rangle_{1} \rightarrow
v^{+}_{1}|vac\rangle_{1}, v^{+}_{1}|vac\rangle_{1} \rightarrow
h^{+}_{1}|vac\rangle_{1},
\end{eqnarray}
and another single-qubit operation on atomic ensemble 2
\begin{eqnarray}
\label{5}h^{+}_{2}|vac\rangle_{2} \rightarrow (h^{+}_{2}-
v^{+}_{2})|vac\rangle_{2}/\sqrt{2},\nonumber\\v^{+}_{2}|vac\rangle_{2}
\rightarrow (h^{+}_{2}+ v^{+}_{2})|vac\rangle_{2}/\sqrt{2}.
\end{eqnarray}
Here, the quantum state of atomic ensembles 1 and 2 becomes
\begin{eqnarray}
\label{6} |\phi\rangle_{12} &=& [h^{+}_{1}(h^{+}_{2}-v^{+}_{2})+
v^{+}_{1}(h^{+}_{2}+v^{+}_{2})]|vac\rangle_{12}/2\nonumber\\&=&
[(h^{+}_{1}\sigma^{2}+v^{+}_{1})
(h^{+}_{2}+v^{+}_{2})]|vac\rangle_{12}/2.
\end{eqnarray}
Obviously the state is a standard bipartite cluster states
($N=2$). The cluster states ($N=2, 3$) can be also generated
without controlled-not transformation \cite{duan1, duan2}.
However, for the generation of the multipartite cluster states,
using the proposals of  Ref. \cite{duan1, duan2} are very hard,
while by the above method with controlled-not transformations is
simply and effective, as shown below. The multipartite cluster
states are also very important in quantum information processing
and quantum computation, so  some of applications have been
proposed \cite{Tame, Zhou, Raussendorf, Nielsen}. Thus the
generation of the multipartite cluster states is necessary.

Out of question, for the generation of arbitrary $N$-particle
cluster state ($N\geq 2$), we can perform the single-qubit
operations and controlled-not transformations to achieve the task
perfectly. Here, we discuss the process in detail. Firstly, we
prepare $N$ atomic ensembles, which are all in the states
$v^{+}_{i}|vac\rangle_{i}$ ($i=1,2\cdots N$). So the state of the
whole system is
\begin{equation}
\label{7} |\phi\rangle_{12\cdots N} = (v^{+}_{1}v^{+}_{2}\cdots
v^{+}_{N})|vac\rangle_{12\cdots N}.
\end{equation}
Secondly, we perform appropriately transformations as the above
process on atomic ensembles 1 and 2 (Eq.(\ref{2})-(\ref{6})),
which lead the initial state to
\begin{eqnarray}
\label{8} |\phi\rangle_{12\cdots N}=
(h^{+}_{1}\sigma^{2}+v^{+}_{1}) (h^{+}_{2}+v^{+}_{2})
\nonumber\\(v^{+}_{3}v^{+}_{4}\cdots
v^{+}_{N})|vac\rangle_{12\cdots N}/2.
\end{eqnarray}
Then, we perform the same transformations on atomic ensembles 2
and 3 as atomic ensembles 1 and 2. We can obtain the result
\begin{eqnarray}
\label{9} |\phi\rangle_{12\cdots N}& =&
(h^{+}_{1}\sigma^{2}+v^{+}_{1})
(h^{+}_{2}\sigma^{3}+v^{+}_{2})\nonumber\\&(h^{+}_{3}&+v^{+}_{3})
(v^{+}_{4}v^{+}_{5}\cdots v^{+}_{N})|vac\rangle_{12\cdots
N}/2\sqrt{2}.
\end{eqnarray}
In a word, if we perform the transformations of Eq.
(\ref{2})-(\ref{6}) on atomic ensembles 1 and 2, then on atomic
ensembles 2 and 3, up to on atomic ensembles $N-1$ and $N$, we
will obtain the perfect multipartite cluster state
\begin{eqnarray}
\label{10} |\phi\rangle_{12\cdots N}& =&
\frac{1}{2^{N/2}}(h^{+}_{1}\sigma^{2}+v^{+}_{1})
(h^{+}_{2}\sigma^{3}+v^{+}_{2})\nonumber\\ &\cdots&
(h^{+}_{N}+v^{+}_{N}) |vac\rangle_{12\cdots N}\nonumber\\&=&
\frac{1}{2^{N/2}}
\otimes^{N}_{i=1}(h^{+}_{i}\sigma^{i+1}+v^{+}_{i})|vac\rangle_{i},
\end{eqnarray}
where $\sigma^{N+1}\equiv1$.

In summary, we have proposed a physical scheme to generate the
cluster states, which are all maximally connected. The cluster
states have some special features: it has a large persistency of
entanglement; it can be regarded as a resource for other
multi-qubit entangled states and so on. So the generation of
cluster state is of great significance in the field of quantum
information. Furthermore, the states have been applied to the
quantum information processing and quantum computation \cite{Tame,
Zhou, Raussendorf, Nielsen}. The cluster states have been
generated in several systems, but the generation via atomic
ensemble is still worthy consideration. Because the schemes based
on atomic ensembles have some special advantages compared with
others , \emph{i.e.} the scheme of atomic ensemble has inherent
fault tolerance function and are robust to realistic noise and
imperfections. At the same time, the theory is simple and
feasible. We can generate the $N$-qubit cluster state simply by
extending the two-qubit case. Our scheme involves Raman-type laser
manipulations, beam splitters, and single-photon detections, and
the requirements of which are all well within  the current
experimental technology.

\begin{acknowledgments}
This work is supported by the Natural Science Foundation of the
Education Department of Anhui Province under Grant No: 2004kj005zd
and Anhui Provincial Natural Science Foundation under Grant No:
03042401 and the Talent Foundation of Anhui University.
\end{acknowledgments}

\end{document}